\documentclass[twocolumn,showpacs,preprintnumbers,amsmath,amssymb]{revtex4}

\usepackage{latexsym}
\usepackage{amsmath}
\usepackage{amssymb}
\usepackage{amsthm}
\usepackage{epsfig}
\usepackage{array}
\usepackage{epsfig}
\usepackage{graphicx}
\usepackage{color}
\usepackage{float}
\usepackage{cancel}

\newtheorem{theorem}{Theorem}

\newtheorem{lemma}[theorem]{Lemma}

\def\<{\langle}
\def\>{\rangle}
\def\ra{\rightarrow}
\def\lbm{ \left[\rule{0pt}{2.1ex}\right. }
\def\rbm{ \left.\rule{0pt}{2.1ex}\right] }

\newcommand{\noshow}[1]{}
\newcommand{\smfrac}[2]{\mbox{$\frac{#1}{#2}$}}

\newcommand{\new}[1]{\textcolor{black}{#1}}

\begin{document}

\title{On the Capacity of Erasure Channel Assisted by Back Classical Communication}

\author{Debbie Leung$^1$\footnote{wcleung@iqc.ca }, Joungkeun Lim$^2$\footnote{canonv@mit.edu} and Peter Shor$^2$\footnote{shor@math.mit.edu}}

\affiliation{$^{1}$ Department of Combinatorics and Optimization, and Institute for Quantum Computing \\ University of Waterloo, 200 University Avenue West Waterloo, Ontario, Canada N2L 3G1 \\ $^{2}$ Department of Mathematics, Massachusetts Institute of Technology \\ 77 Massachusetts Avenue, Cambridge, MA 02139, USA}

\date{\today}


\begin{abstract}
We present a communication protocol for the erasure channel assisted by backward classical communication, which achieves a significantly better rate than the best prior result. In addition, we prove an upper bound for the capacity of the channel. The upper bound is smaller than the capacity of the erasure channel when it is assisted by two-way classical communication. Thus, we prove the separation between quantum capacities assisted by backward classical communication and two-way classical communication.
\end{abstract}


\maketitle

In quantum information theory, a {\it capacity} $Q(\chi)$ of a channel $\chi$ is a theoretical maximum of the rate $m/n$ that is achievable by some communication protocol that sends $m$-qubit information with $n$ uses of the channel, where $n$ tends to infinity.
The above definition of $Q$ is defined without auxiliary resources, and additional free classical communication may increase the capacity. We use $Q$, $Q_1,Q_{\rm B}$, and $Q_2$ to denote the quantum capacities of a quantum channel when unassisted, assisted by unlimited forward, backward, and two-way classical communication, respectively. It was proved that classical forward communication alone does not increase the quantum capacity of any channel; in other words $Q(\chi)=Q_1(\chi)$ for all channels $\chi$ \cite{BDSW}. In contrast, $Q_2$ is greater than $Q$ for some channels \cite{BDSW}. $Q_{\rm B}$ is also known to be greater than $Q$ for some channels \cite{BDS}, but it has been an open question whether $Q_{\rm B}(\chi) = Q_2(\chi)$ for all $\chi$.

We study the capacities of the quantum erasure channel, which was first introduced in \cite{GBP}. The quantum erasure channel of erasure probability $p$, denoted by ${\cal N}_p$, replaces the incoming qubit, with probability $p$, with an ``erasure state'' $|2\>$ orthogonal to both $|0\>$ and $|1\>$, thereby both erasing the qubit and informing the receiver that it has been erased. In an equivalent formulation, called the isometric extension, the channel exchanges the incoming qubit with the environmental system in state $|2\>$ with probability $p$. It was shown in \cite{BDS} that the quantum capacities $Q, Q_1$, and $Q_2$ for ${\cal N}_p$ are given by
\begin{equation*}
\begin{split}
&Q({\cal N}_p) = Q_1({\cal N}_p) = \max \{0,1-2p\} \text{ and }\\
&Q_2({\cal N}_p) = 1-p.
\end{split}
\end{equation*}
However, until the current investigation, little has been known about $Q_{\rm B}({\cal N}_p)$ except for two lower bounds that follow straightforwardly from 1-way hashing \cite{BDSW} and teleportation \cite{BBCJPW} and an upper bound given by $Q_2({\cal N}_p)$ as
\begin{equation} \label{previous result}
\begin{split}
&Q_{\rm B}({\cal N}_p) \geq 1-2p, \text{ if } p \leq 2/5,\\
&Q_{\rm B}({\cal N}_p) \geq (1-p)/3, \text{ if } p \geq 2/5, \text{ and }\\
&Q_{\rm B} ({\cal N}_p)\leq Q_2({\cal N}_p) = 1-p.
\end{split}
\end{equation}

In this Letter, we present an efficient communication protocol that achieves a better lower bound of $Q_{\rm B}({\cal N}_p)$, and we prove a new upper bound of $Q_{\rm B}({\cal N}_p)$. With this upper bound, we show that $Q_{\rm B}({\cal N}_p) < Q_2({\cal N}_p)$ for all $p$ and resolve the previously open question.

{\bf Preliminaries and Notations--}
Recall the definition of {\it von Neumann entropy} $H(A)=H(\psi^A)=-tr(\psi^A \log \psi^A)$, where $\psi^A$ is the density operator for system $A$. The {\it quantum mutual information} and {\it coherent information} are defined as
\begin{align*}
&I(A;B)=H(A)+H(B)-H(AB),
\text{ and }\\
&I(A \rangle B) = H(B)-H(AB).
\end{align*}
The statements in the following lemma will be used in the proof of a theorem in the later section.

\begin{lemma}\label{lemma}
For disjoint systems A, B, and C,\\
(i)~$I(AB;C)-I(B;C)\leq I(A;BC).$\\
(ii)~$I(A \rangle B) \leq I(A \rangle BC).$\\
(iii)~$I(A \rangle C)+I(B \rangle C) \leq I(AB \rangle C).$\\
(iv)~$I(A \rangle BC) - I(A \rangle B) \leq 2H(CE)$, 
where E is any subset of B.
\end{lemma}

\begin{proof}
Subadditivity and strong subadditivity inequalities \cite{NC} easily give $(i),(ii),(iii),$
\begin{align*}
&H(CDE) \leq H(D)+H(CE),\\
&H(AD) \leq H(CE) + H(ADCE), \text { and }\\
&H(D)+H(ADE) \leq H(AD) + H(DE),
\end{align*}
for $E\subset B$ and $D=B/E$.
Adding these three inequalities yields $(iv)$.
\end{proof}

We consider only near-perfect communication protocols that produce, with high probability, output states of high {\it fidelity} with the input states. The fidelity of states $\rho_{in}$ and $\rho_{out}$ is defined to be
\begin{equation*}
F(\rho_{in},\rho_{out})\equiv \text{\rm tr} \sqrt { \rho_{in}^{1/2} \rho_{out} \rho_{in}^{1/2}}.
\end{equation*}
From now on, we call the sender, the receiver, and the environment Alice, Bob, and Eve.

{\bf Communication Protocol--}
we derive an improved lower bound for $Q_{\rm B}({\cal N}_p)$ by providing a communication protocol. The protocol combines two subprotocols that utilize {\it coherent teleportation} introduced in \cite{H}.

{\it Coherent Teleportation--}
Given an unknown qubit state $|\psi \rangle = a|0\>+b|1\>$ in system $M$ and an {\em ebit} (sometimes called an EPR pair or Bell state) $|\Phi\>_{AB} = \smfrac{1}{\sqrt{2}} (|00\>+|11\>)$ between Alice and Bob, Alice can transmit $|\psi\>$ to Bob by teleportation \cite{BBCJPW}.
In the original teleportation protocol, the change of basis takes the initial state $|\psi\>_M |\Phi\>_{AB}$ to
\begin{equation}\label{state1}
        \smfrac{1}{2} \sum_{ij} |ij\>_{MA} \; X^i Z^j |\psi\>_B  \,.
\end{equation}
Reference \cite{H} proposes a {\em coherent} variant of teleportation in which Alice does not measure $|ij\>_{MA}$ but instead, coherently copies $|ij\>_{MA}$ to two ancillary systems ${C_1 C_2}$ and transmits them coherently to Bob.
Mathematically, Alice and Bob share the joint state
%
%
$\smfrac{1}{2} \sum_{ij} |ij\>_{MA} \; |ij\>_{C_1 C_2} \; X^i Z^j |\psi\>_B$.
After receiving $C_1 C_2$, Bob can apply a control-$X$ from $C_1$ to $B$ and then a control-$Z$ from $C_2$ to $B$.  Alice and Bob then share the state
%
$\smfrac{1}{2} \sum_{ij} |ij\>_{MA} \; |ij\>_{C_1 C_2} \; |\psi\>_B$,
with $|\psi\>$ transmitted {\em and} two ebits shared between Alice and Bob.

{\it First Subprotocol--}
Suppose Alice and Bob already share an ebit, and Alice teleports $|\psi\>$ to Bob by attempting to use the erasure channel for coherent classical communication of each of $|i\>_{C_1}$ and $|j\>_{C_2}$ (see previous subsection on coherent teleportation). Bob tells Alice whether the communication is erased or not.  If so, Alice copies and sends it again until Bob receives it.
Note that the transmission is coherent if it is not erased in the first trial.
If $i$ and $j$ are erased $k$ and $l$ times before they are sent successfully, the state becomes (after Bob's controlled-$X$ and $Z$)
\begin{align*}
&\smfrac 1 2 \sum_{ij} |ij\rangle_A |i\rangle_E^{\otimes k} |j\rangle_E^{\otimes l} |ij\rangle_B |\psi\rangle_B \\
&\sim
|\Gamma \rangle _{ABE} ^{\otimes (1_k{+}1_l)}
|\Phi \rangle_{AB}^{\otimes (2{-}1_k{-}1_l)}  |\psi\rangle_B,
\end{align*}
where $1_k = 0$ if $k=0$ and $1_k = 1$ if $k>0$ and similarly for
$1_l$, $|\Gamma\>=\smfrac{1}{\sqrt{2}} (|000\>+|111\>)$, and $\sim$
denotes equivalence up to a unitary transformation on $E$.

Since the success probability of each transmission is $1-p$, Alice tries $\smfrac 1 {1-p}$ times on average to send each register $i$ and $j$. Hence she transmits $\smfrac 2 {1-p}$ qubits through the channel. Both $1_k$ and $1_l$ have expectation $p$. 
In {\it asymptotic resource inequality} [6],
\begin{equation} \label{le1}
\smfrac{2}{1{-}p} ~{\cal N}_p + \Phi_{AB} ~\geq~ 1~ \text{Qbit} + 2(1{-}p)~\Phi_{AB} + 2p~\Gamma_{ABE},
\end{equation}
where resources on the left-hand side simulate those on the right, ${\cal N}_p$ denotes one use of the erasure channel, and Qbit denotes one use of the noiseless qubit channel. We have used $\Phi$ and $\Gamma$ as shorthand for $|\Phi\>\<\Phi|$ and $|\Gamma\>\<\Gamma|$.
With free back classical communication, one use of ${\cal N}_p$ can prepare one ebit with probability $1-p$.  Hence,
\begin{equation} \label{le2}
1~ {\cal N}_p \geq (1{-}p)~ \Phi_{AB}.
\end{equation}
We combine equations (\ref{le1}) and (\ref{le2}) to get
\begin{align*}
& 1~ {\cal N}_p \geq \smfrac {1-p} 2 ~\text{ Qbit, if } p \leq 1/2 \text{ and }\\
& 1~ {\cal N}_p \geq \smfrac {1-p}{1+2p}~ \text{ Qbit, if } p \geq 1/2.
\end{align*}
Hence, the rate of the first subprotocol is
\begin{align*}
&\smfrac {1-p} 2  \text{ , if } p\leq 1/2 \text{ and }\\
&\smfrac {1-p}{1+2p}  \text{ , if } p\geq 1/2.
\end{align*}

{\it Second Subprotocol--}
This method only differs from the previous subprotocol in that $|ij\>$ will be sent using a coherent version of superdense coding.
More specifically, in this case, Alice and Bob first share an ebit $|\Phi\>_{C_1 C_2}$ where $C_1$ belongs to Alice and $C_2$ belongs to Bob.  After the change of basis (see equation (\ref{state1})), Alice applies control-$X$ from $M$ to $C_1$ and control-$Z$ from $A$ to $C_1$, resulting in the joint state
\begin{equation*}
        \smfrac{1}{2} \sum_{ij} |ij\>_{MA} \; |\Phi_{ij}\>_{C_1 C_2} \;
        X^i Z^j |\psi\>_B  \,,
\end{equation*}
and sends $C_1$ to Bob using the erasure channel. The states $|\Phi_{ij}\>=X^i Z^j |\Phi\>$ are orthogonal (they form the {\it Bell basis}) \cite{NC}. In case of erasure, Bob and Eve share $|\Phi_{ij}\>_{C_1 C_2}$ and Alice and Bob will take another ebit and repeat the superdense coding procedure, until Bob receives the transmission (call the two-qubit system in his possession $D_1D_2$). Then, Bob applies the transformation $|\Phi_{ij}\>_{D_1D_2} \ra |ij\>_{D_1D_2}$ and coherently reverts the $X^iZ^j$ not only in $X^i Z^j |\psi\>_B$ but also in all the $|\Phi_{ij}\>$ he shares with Eve (by acting only on his halves), so that the final state becomes
\begin{equation*}
\smfrac 1 2 \sum_{ij} |ij\rangle_{MA} |ij\>_{D_1D_2} \;
|\Phi\rangle_{EB}^{\otimes k} \; |\psi\rangle_B.
\end{equation*}
where $k$ again denotes the number of erasures before the successful transmission.  In this method, Alice and Bob {\em always} share $2$ ebits at the end.

Once again, Alice needs to apply superdense coding $\smfrac{1}{1{-}p}$ times on average. This gives the asymptotic resource inequality,
\begin{align*}
&\Phi_{AB} + \smfrac{1}{1{-}p} \lbm {\cal N}_p + \Phi_{AB} \rbm\\
&~\geq~ 1 \text{ Qbit } + 2 ~\Phi_{AB} + (\smfrac{1}{1{-}p} - 1) ~\Phi_{BE} \,.
\end{align*}
Note that the above consumes more ebits than it produces for all $p$; thus, we use equation (\ref{le2}) to supply the needed ebits, and obtain
\begin{equation*}
1~{\cal N}_p \geq (1-p)^2 \text{ Qbit}.
\end{equation*}
Hence the rate of the second subprotocol is $(1-p)^2$.

{\it Rate of Communication Protocol--}
Applying the two protocols selectively, the rate of the protocol is
\begin{equation}\label{current lowerbound}
\begin{split}
(1-p)^2 &\text{ , if } p\leq 1/2 \text{ and }\\
\smfrac {1-p}{1+2p}  &\text{ , if } p\geq 1/2.
\end{split}
\end{equation}

{\bf Upper Bound for the Capacity--}
The purpose of this section is to prove that $Q_{\rm B}({\cal N}_p) \leq \smfrac {1-p}{1+p}$. By the definition of the capacity, for each $n$, there is a protocol ${\cal P}_n$ that uses back classical communication and ${\cal N}_p$ at most $n$ times and transmits $n(Q_{\rm B}({\cal N}_p) - \delta_n)$ qubits from Alice to Bob with fidelity at least $1-\epsilon_n$ and probability at least $1-\epsilon_n$, where $\epsilon_n,\delta_n \ra 0$ as $n \ra \infty$.

Our strategy to show the upper bound is as follows.  We consider any protocol that transmits $m$ qubits with $n$ uses of the channel.  In particular, such protocol must be able to transmit $m$ halves of ebits shared between Alice and a reference system $R$ \cite{BKN}, without entangling Eve and $R$ (or else the transmission to Bob will be noisy).  This translates to bounds on quantum mutual information between Bob, Eve, and $R$ that will be contradicted if $m/n$ is larger than our stated upper bound.

If Alice transmits her halves of the ebits shared with $R$ directly through the channel, any loss to Eve can never be recovered.  Thus, Alice has to transmit quantum states whose potential entanglement with $R$ can be materialized or nullified depending on Bob's back communication and Alice's future transmissions.  The finalizing or nullifying process requires further uses of the channel, giving an upper bound to the capacity.

To quantify the above idea, denote by $S_1, S_2, \cdots S_n$ the
qubits transmitted by Alice through the channel.  Each $S_i$ is
delivered to Bob with probability $1-p$ or lost to Eve with
probability $p$.  Let $\mathcal{B}=\{i|S_i \text{ sent to Bob} \}$ and
$\mathcal{E}=\{i|S_i \text{ sent to Eve}\}$ be the index sets of
qubits delivered to Bob and Eve.
We define $E_i=\bigcup _{1\leq j \leq i \,, \; j \in \mathcal{E}} S_j$
to be Eve's system after the $i$th transmission.  For Bob, the most
general procedure after each transmission is an isometry followed by a
measurement.  By double-block coding and by extending Theorem 10 in
\cite{winter}, any such measurement can be approximated by a von
Neumann measurement on part of Bob's system (turning the measured
qubits into classical data).
Let $\tilde{B}_i$ be Bob's quantum system immediately after the $i$th
channel use, and $B_i$ be his quantum system after his measurement and
classical feedback to Alice.
Thus $\tilde{B}_i=B_{i-1}\cup S_i$ if $S_i$ is delivered to Bob, and
$\tilde{B}_i=B_{i-1}$ if $S_i$ is lost to Eve.
Suppose a total of $c$ qubits are measured by Bob in the
protocol.
%
After the final decoding
operation, Bob produces an $m$-qubit system $B^{(1)}$ that is almost
maximally entangled with the system $R$. We denote the rest of Bob's
quantum 
system by $B^{(2)}$. 


In the following theorem, $I(S_i; B_iR)$ is the amount of mutual
information carried by each transmission $S_i$.  
For the rest of the
paper, information theoretical quantities are evaluated on the states
that are held at the corresponding stages of the protocol.
Part $(i)$ of the theorem states that a sufficient amount of mutual
information (2$m$ for $m$ ebits) has to be delivered to Bob. Part
$(ii)$ states that the more mutual information is lost to Eve, the
more transmissions are needed to nullify the lost information.

\begin{theorem}\label{theorem1}
If the fidelity between the input and output states is at least $1-\epsilon_n$, then \\
(i)~$\sum_{i\in \mathcal{B}} \; I(S_i ; B_{i-1}R) \geq 2m-2(2\sqrt 2 m \sqrt{\epsilon_n}+1).$\\
(ii)~$\sum _{i\in \mathcal{E}} \; I(S_i ; B_{i-1}R) \leq n-m+4(2 \sqrt 2 m \sqrt{\epsilon_n}+1).$
\end{theorem}

\begin{proof}

$(i)$ For each $i\in \mathcal{B}$, apply part $(i)$ of lemma \ref{lemma} on the systems $S_i, B_{i-1}$, and $R$ to obtain
\begin{align*}
I(B_i;R)-I(B_{i-1};R) & \leq I(B_{i-1}S_i;R)-I(B_{i-1};R)\\
&\leq I(S_i ; B_{i-1}R).
\end{align*}
Thus,
\begin{align*}
\sum_{i\in \mathcal{B}} I(S_i ; B_{i-1}R) 
&\geq \sum _{i\in \mathcal{B}} (I(B_i;R)-I(B_{i-1};R)) \\
&=I(B_n;R) =I(B^{(1)}B^{(2)};R) \\ 
&\geq I(B^{(1)};R)\\
&=H(B^{(1)})+H(R)-H(B^{(1)}R)\\
&\geq 2(H(R)-H(B^{(1)}R)).
\end{align*}

Note that the fidelity between the state $\mu$ in $B^{(1)}R$ and \new{$\Phi^{\otimes m}$} is at least $1{-}\epsilon_n$. Let $D=\smfrac 1 2 {\rm tr}|\mu - \Phi ^{\otimes m}|$ be the {\it trace distance} \cite{NC} between $\mu$ and $\Phi^{\otimes m}$. By page 415 of \cite{NC}, 
$$D\leq \sqrt {1-F(\mu, \Phi ^{\otimes m})^2} \leq \sqrt {2\epsilon_n}.$$
By Fannes' inequality \cite{NC},
\begin{align*}
H(B^{(1)} R)
&=|H(\mu)-H(\Phi^{\otimes m})| 
\leq 2Dm-2D\log (2D)\\
&\leq 2\sqrt 2m \sqrt {\epsilon_n} +1.
\end{align*}

$(ii)$ Using $2$, $3$, and $4$ to denote the use of parts
$(ii),(iii),$ and $(iv)$ of lemma \ref{lemma} respectively, we have
\begin{align*}
& \sum _{i\in \mathcal{E}} I(S_i \rangle B_{i-1}R)
\stackrel{2}{\leq} 
c + \sum _{i\in \mathcal{E}} I(S_i \rangle B_nR) \\
&\stackrel{3}{\leq} c + I(\bigcup _{i\in \mathcal{E}}S_i \rangle B_nR)
= c + I(E_n \rangle B^{(1)}B^{(2)}R)\\
&\stackrel{4}{\leq} c + I(E_n \rangle B^{(1)}B^{(2)}) + 2H(B^{(1)}R)\\
&\stackrel{3}{\leq} c + I(E_nR \rangle B^{(1)}B^{(2)}) - I(R \rangle B^{(1)}B^{(2)}) + 2H(B^{(1)}R)\\
&= c + I(E_nR \rangle B_n) - I(R \rangle B^{(1)}B^{(2)}) + 2H(B^{(1)}R)
\end{align*}
where the equalities use the fact that Bob's decoding is isometric.
$I(E_nR \rangle B_n)$ is upper bounded by 
$n-|\mathcal {E}|-c$. 
$I(R \rangle B^{(1)}B^{(2)})$ is lower bounded as
\begin{align*}
I(R \rangle B^{(1)}B^{(2)}) & \stackrel{2}{\geq} I(R \rangle B^{(1)}) 
\stackrel{4}{\geq} I(R \rangle B^{(1)}T)-2H(T)\\
&= m-  2H(B^{(1)}R)
\end{align*}
where $T$ purifies $B^{(1)}R$. Putting together the two previous sets of inequalities,
\begin{equation*}
\sum _{i\in \mathcal{E}} I(S_i \rangle B_{i-1}R)
\leq n-|\mathcal {E}| -m +4(2\sqrt 2 m \sqrt {\epsilon_n}+1).
\end{equation*}
Hence,
\begin{align*}
\sum _{i\in \mathcal{E}} I(S_i ; B_{i-1}R) &= \sum _{i\in \mathcal{E}} (H(S_i) + I(S_i \rangle B_{i-1}R))\\
&\leq \sum _{i\in \mathcal{E}} (1+ I(S_i \rangle B_{i-1}R))
\noshow{&\leq n -m +4(2\sqrt 2 m \sqrt{\epsilon_n}+1).}
\end{align*}
$\hspace{97pt}\leq n -m +4(2\sqrt 2 m \sqrt{\epsilon_n}+1).$
\end{proof}
\new{Since Alice cannot predict whether Bob or Eve will receive the next transmission and a certain fraction of the transmission are lost to Eve, the same fraction of mutual information has to be lost to Eve. Combined with the theorem, the argument gives an upper bound of $Q_B({\cal N}_p)$.}  \new{To prove this rigorously,} consider the following random variable.
$$X_i = \left\{\begin{array}{ll}
\smfrac p 2 ~I(S_i;B_{i-1}R) &\text{ if } S_i \text{ is delivered to Bob}\\
\smfrac {-(1-p)} 2 ~I(S_i;B_{i-1}R) &\text{ if } S_i\text{ is lost to Eve}\\
\end{array} \right.
$$
Then $|X_i|\leq 1$ and $E(X_i)=0$. Note that the $X_i$'s may not be independent variables. Let $Y_i=\sum _{j=1} ^i X_j$ and $Y_0=0$. Then $Y_0,Y_1, \cdots, Y_n$ is a {\it martingale} \new{\cite{AS}} with $|Y_{i+1}-Y_i|\leq 1$. If the fidelity between the input and output states is at least $1-\epsilon_n$, then from theorem \ref{theorem1}
\begin{align*}
Y_n
&=\smfrac p 2 \sum _{i\in \mathcal{B}} I(S_i;B_{i-1}R) -\smfrac {(1-p)} 2 \sum _{i\in \mathcal{E}} I(S_i;B_{i-1}R)\\
&\geq \smfrac {(1+p)} 2 m- \smfrac {(1-p)} 2 n- (2-p)(2\sqrt 2 m \sqrt{\epsilon_n}+1).
\end{align*}
Assume by contradiction that $Q_{\rm B}({\cal N}_p) > \smfrac {1-p}{1+p}$.  Then, for sufficiently large $n$, $\smfrac{m}{n} \geq \smfrac{1-p}{1+p} + 4k$ for some $k>0$.  The above expression for $Y_n$, which holds with probability at least $1-\epsilon_n$, will exceed $kn$. Therefore $\lim_{n\rightarrow \infty}\text{Pr} [\; |Y_n|\geq kn ]=1.$
\noshow{
\begin{equation*}
\lim_{n\rightarrow \infty}\text{Pr} [\; |Y_n|\geq kn ]=1.
\end{equation*}
}

However, Azuma's inequality \cite{AS} applied to martingale $Y_i$ gives $\Pr[|Y_n|\geq kn] \leq e^{-\smfrac {k^2} 2 n}$, and $\lim_{n\rightarrow\infty}\text{Pr} [\; |Y_n|\geq kn] =0,$
\noshow{
\begin{equation*}
\lim_{n\rightarrow\infty}\text{Pr} [\; |Y_n|\geq kn] =0,
\end{equation*}
}
which is a contradiction. Hence,
\begin{equation}\label{current upperbound}
Q_{\rm B}({\cal N}_p) \leq \smfrac {1-p}{1+p}.
\end{equation}

{\bf Discussion--}
We summarize the previous and our new results in Figure 1.
The lighter region is the previous undetermined area of $Q_{\rm B}({\cal N}_p)$, given by the previous lower and upper bounds in equation (\ref{previous result}).
The darker region is the new undetermined area of $Q_{\rm B}({\cal N}_p)$ due to our lower and upper bounds in equations (\ref{current lowerbound}) and (\ref{current upperbound}), which are significantly improved over previous results.
Since our upper bound of $Q_{\rm B}({\cal N}_p)$ is strictly less than
$Q_2({\cal N}_p)$, we prove the separation between $Q_{\rm B}$ and
$Q_2$ answering the long-standing question raised in \cite{BDS}.
\begin{figure}
\epsfig{file=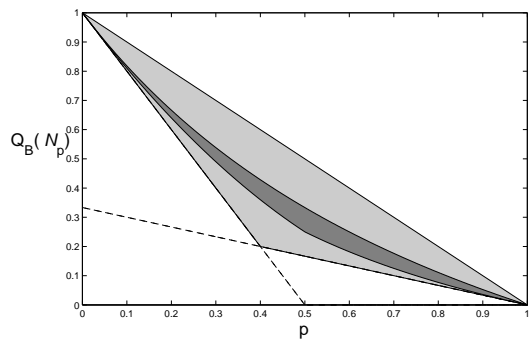,width=3in}
\caption{Undetermined area of $Q_{\rm B}({\cal N}_p)$}
\label{figure1}
\end{figure}

%

We thank Andrzej Grudka and Michal Horodecki for pointing out an
important mistake in an earlier manuscript, and for suggesting a
solution that also substantially simplifies the proof.
This research was partially supported by the W. M. Keck Foundation Center for Extreme Quantum Information Theory. P.W.S. and J.L. would like to thank the National Science Foundation for support through grant CCF-0431787.
J.L. thanks SLSF for support. D.L. thanks NSERC, CRC, CFI, ORF, MITACS, ARO, and CIFAR for support.


\begin{thebibliography}{8}
\expandafter\ifx\csname natexlab\endcsname\relax\def\natexlab#1{#1}\fi
\expandafter\ifx\csname bibnamefont\endcsname\relax
  \def\bibnamefont#1{#1}\fi
\expandafter\ifx\csname bibfnamefont\endcsname\relax
  \def\bibfnamefont#1{#1}\fi
\expandafter\ifx\csname citenamefont\endcsname\relax
  \def\citenamefont#1{#1}\fi
\expandafter\ifx\csname url\endcsname\relax
  \def\url#1{\texttt{#1}}\fi
\expandafter\ifx\csname urlprefix\endcsname\relax\def\urlprefix{URL }\fi
\providecommand{\bibinfo}[2]{#2}
\providecommand{\eprint}[2][]{\url{#2}}

\bibitem[{\citenamefont{Bennett et~al.}(1996)\citenamefont{Bennett, DiVincenzo,
  Smolin, and Wootters}}]{BDSW}
\bibinfo{author}{\bibfnamefont{C.~H.} \bibnamefont{Bennett}},
  \bibinfo{author}{\bibfnamefont{D.~P.} \bibnamefont{DiVincenzo}},
  \bibinfo{author}{\bibfnamefont{J.~A.} \bibnamefont{Smolin}},
  \bibnamefont{and} \bibinfo{author}{\bibfnamefont{W.~K.}
  \bibnamefont{Wootters}}, \bibinfo{journal}{Phys. Rev. A.}
  \textbf{\bibinfo{volume}{54}}, \bibinfo{pages}{3824} (\bibinfo{year}{1996}).

\bibitem[{\citenamefont{Bennett et~al.}(1997)\citenamefont{Bennett, DiVincenzo,
  and Smolin}}]{BDS}
\bibinfo{author}{\bibfnamefont{C.~H.} \bibnamefont{Bennett}},
  \bibinfo{author}{\bibfnamefont{D.~P.} \bibnamefont{DiVincenzo}},
  \bibnamefont{and} \bibinfo{author}{\bibfnamefont{J.~A.}
  \bibnamefont{Smolin}}, \bibinfo{journal}{Phys. Rev. Lett.}
  \textbf{\bibinfo{volume}{78}}, \bibinfo{pages}{3217} (\bibinfo{year}{1997}).

\bibitem[{\citenamefont{Grassl et~al.}(1997)\citenamefont{Grassl, Beth, and
  Pellizzari}}]{GBP}
\bibinfo{author}{\bibfnamefont{M.}~\bibnamefont{Grassl}},
  \bibinfo{author}{\bibfnamefont{T.}~\bibnamefont{Beth}}, \bibnamefont{and}
  \bibinfo{author}{\bibfnamefont{T.}~\bibnamefont{Pellizzari}},
  \bibinfo{journal}{Phys. Rev. A.} \textbf{\bibinfo{volume}{56}},
  \bibinfo{pages}{33} (\bibinfo{year}{1997}).

\bibitem[{\citenamefont{Bennett et~al.}(1993)\citenamefont{Bennett, Brassard,
  Cr\'{e}peau, Jozsa, Peres, and Wootters}}]{BBCJPW}
\bibinfo{author}{\bibfnamefont{C.~H.} \bibnamefont{Bennett}},
  \bibinfo{author}{\bibfnamefont{G.}~\bibnamefont{Brassard}},
  \bibinfo{author}{\bibfnamefont{C.}~\bibnamefont{Cr\'{e}peau}},
  \bibinfo{author}{\bibfnamefont{R.}~\bibnamefont{Jozsa}},
  \bibinfo{author}{\bibfnamefont{A.}~\bibnamefont{Peres}}, \bibnamefont{and}
  \bibinfo{author}{\bibfnamefont{W.~K.} \bibnamefont{Wootters}},
  \bibinfo{journal}{Phys. Rev. Lett.} \textbf{\bibinfo{volume}{70}},
  \bibinfo{pages}{1895} (\bibinfo{year}{1993}).

\bibitem[{\citenamefont{M.A.Nielsen and I.L.Chuang}(2000)}]{NC}
\bibinfo{author}{\bibnamefont{M.A.Nielsen}} \bibnamefont{and}
  \bibinfo{author}{\bibnamefont{I.L.Chuang}}, \emph{\bibinfo{title}{Quantum
  Computation and Quantum Information}} (\bibinfo{publisher}{Cambridge
  University Press}, \bibinfo{year}{2000}).

\bibitem[{\citenamefont{Harrow}(2004)}]{H}
\bibinfo{author}{\bibfnamefont{A.~W.} \bibnamefont{Harrow}},
  \bibinfo{journal}{Phys. Rev. Lett.} \textbf{\bibinfo{volume}{92}},
  \bibinfo{pages}{097902} (\bibinfo{year}{2004}).

\bibitem[{\citenamefont{Barnum et~al.}(2000)\citenamefont{Barnum, Knill, and
  M.A.Nielsen}}]{BKN}
\bibinfo{author}{\bibfnamefont{H.}~\bibnamefont{Barnum}},
  \bibinfo{author}{\bibfnamefont{E.}~\bibnamefont{Knill}}, \bibnamefont{and}
  \bibinfo{author}{\bibnamefont{M.A.Nielsen}}, \bibinfo{journal}{IEEE Trans.
  Inf. Theory} \textbf{\bibinfo{volume}{46}}, 
  \bibinfo{pages}{1317}
  (\bibinfo{year}{2000}).

\bibitem{winter}
\bibinfo{author}{\bibfnamefont{A.}~\bibnamefont{Winter}},
  \bibinfo{journal}{Comm.\ Math.\ Phys.} 
  \textbf{\bibinfo{volume}{244}}, \bibinfo{pages}{157} (\bibinfo{year}{2004}).



\bibitem[{\citenamefont{Alon and Spencer}(2000)}]{AS}
\bibinfo{author}{\bibfnamefont{N.}~\bibnamefont{Alon}} \bibnamefont{and}
  \bibinfo{author}{\bibfnamefont{J.~H.} \bibnamefont{Spencer}},
  \emph{\bibinfo{title}{The probabilistic method}}
  (\bibinfo{publisher}{Wiley-Interscience, New York}, \bibinfo{year}{2000}).

\end{thebibliography}
\end{document}